\documentclass[aps,prl,preprint,superscriptaddress,floatfix,longbibliography]{revtex4-1}
\usepackage{amsmath}
\usepackage{amsfonts} 
\usepackage{graphicx}
\usepackage{verbatim}
\usepackage{natbib}
                                                                    
\setcitestyle{super}

\begin{document}

\title{Strong light-matter coupling in MoS$_2$}

\author{Patryk Kusch}
\author{Niclas S. Mueller}
\author{Martin Hartmann}
\author{Stephanie Reich}

\affiliation{Freie Universit\"at Berlin, Department of Physics, Arnimallee 14, 14195 Berlin}

\date{\today}

\begin{abstract}
Polariton-based  devices require materials where light-matter coupling under ambient conditions exceeds  losses, but our current selection of such materials is limited. Here we measured the dispersion of polaritons formed by the $A$ and $B$ excitons in thin MoS$_2$ slabs by imaging their optical near fields. We combined fully tunable laser excitation in the visible with a scattering near-field optical microscope to excite polaritons and image their optical near fields. We obtained the properties of bulk MoS$_2$ from fits to the slab dispersion. The in-plane excitons are in the strong regime of light-matter coupling with a coupling strength ($40-100\,$meV) that exceeds their losses by at least a factor of two. The coupling becomes comparable to the exciton binding energy, which is known as very strong coupling. MoS$_2$ and other transition metal dichalcogenides are excellent materials for future polariton devices.
\end{abstract}

\maketitle

\section{Introduction}

Exciton polaritons are mixed states of light and matter that form if the interaction strength between the exciton as a material excitation and  photons exceeds the  losses in the material.\cite{Baranov2018} These quasi-particles are promising to transmit and convert information as they  have long propagation length and mix the light and matter degrees of freedom. Polaritons have been studied profoundly in matter-filled photonic cavities and artificial nanoscale systems.\cite{Baranov2018,Kockum2019} Observing them in three-dimensional (3D) materials, however, proved challenging, because the oscillator strengths of 3D excitons are often weak and their binding energies are smaller than the thermal energy at room temperature.\cite{CardonaBuch,Adreani1995} For example, the polariton coupling energy of the exciton in GaAs is $g=7.8\,$meV.\cite{Adreani1995} Measuring GaAs polaritons requires high quality  crystals and cryogenic temperatures. Similar values are found in many other bulk semiconductors.\cite{Sanvitto2016} A notable exception are wide band-gap semiconductors like ZnO and GaN, which makes them promising materials for polariton devices like ultra-low threshold lasers.\cite{Sun2008,Kang2019,Sanvitto2016} The polariton coupling strength in, e.g., ZnO reaches 60\,meV for the $A$ and 140\,meV for the $B$ exciton that 
both have energies $\approx 3.31\,$eV (Ref.~\cite{Sun2008}). The high exciton energies in ZnO and similar semiconductors, however, restrict such devices to the ultraviolett energy range. In addition, wide band-gap semiconductors have low dielectric constants limiting their ability to confine and guide light.

Transition metal dichalcogenides (TMDCs) are layered semiconductors with band gaps in the visible and near-infrared energy range. TMDCs have  high exciton binding energies (50-200\,meV), a high oscillator strength and large dielectric constants.\cite{Wang2018,Munkhbat2019,Hu2019,Neville1976,Ermolaev2020} They are covalently bound within the layers with a chemical composition MX$_2$, where M=Mo, W and X=S, Se.\cite{Wang2018} The layers are held together by van-der-Waals forces, which allows to exfoliate TMDCs down to the limit of a single monolayer. TMDCs have two in-plane polarized excitons, the $A$ and $B$ exciton that both arise from transitions at the $K$ point of the Brillouin zone.\cite{Mak2010,Fortin1975,Neville1976,Wang2018,Schneider2018}  When the TMDCs are inserted into photonic cavities, these excitons form polaritons with cavity photons.\cite{Schneider2018,Hu2019,Munkhbat2019}  First studies of light-matter coupling in TMDCs focused on monolayers in metallic or dielectric cavities.\cite{Liu2014,Hu2019,Flatten2016} The polariton dispersion of an  MoS$_2$ monolayer was measured by scanning the incident angle of the light in a cavity setup.\cite{Liu2014} The system had two distinct polariton branches that were separated by a gap with the minimum energy $\Omega_r=2 g=46\,$meV (Rabi splitting). Attempts to similarly measure the dispersion in free-standing flakes ($10-100\,$nm thickness) by reflectance were hindered by the broad spectral features and the large index of refraction that restricts the accessible propagation direction to be within $20^\circ$ of the plane normal.\cite{Munkhbat2019} Estimates of the coupling strength from the dispersion around $\Gamma$ gave values of $30-60\,$meV for the $A$ and $B$ exciton alike. Hu~\textit{et al.}\cite{Hu2017} introduced scattering scanning near field optical microscopy (s-SNOM) to access the polariton dispersion of TMDCs in the near infrared. They imaged the dispersion of MoSe$_2$ around the energy of the $A$ exciton ($1.55\,$eV) and found $\Omega_r\approx 100\,$meV. The light-matter coupling strength in monolayers and thin slabs of TMDCs ($g\approx 20-50\,$meV) appears much higher than in classical semiconductors.
The coupling strength in bulk TMDCs may be argued to be even larger, because only a fraction of the electromagnetic field overlaps with the material in thin samples.\cite{Savona1995} 

The second parameter that controls the regime of strong light-matter coupling is the decay rate of the coupling states.\cite{Baranov2018}
Specifically, strong light-matter coupling is reached if $g$ is larger than the homogeneous broadening. Polariton formation and strong coupling appear robust against inhomogeneous broadening as long as the inhomogeneous broadening remains smaller than the Rabi splitting.\cite{Manceau2017}
The typical exciton linewidths of TMDC monolayers ($50\,$meV, Refs.~\cite{Wang2018,Mak2010,Saigal2016,Carvalho2015}) are caused by inhomogeneous broadening as shown exemplary for a WSe$_2$ monolayer where the homogeneous width (2\,meV) was an order of magnitude smaller than the peak width (50\,meV) of the $A$ exciton at low temperatures.\cite{Moody2015,Cadiz2017} The total decay times of the $A$ and $B$ excitons in bulk MoS$_2$ were measured and calculated as $\approx 1-2\,$ns ($0.5\,\mu$eV).\cite{Shi2013,Palummo2015,Cadiz2017,Cha2016,Wang2018PCC}  Reaching the regime of strong coupling in TMDCs appears within easy reach when considering the narrow intrinsic width of the exciton lines. On the other hand, the inhomogeneous broadening is strong and may prevent the observation of the effect. Given the uncertainties in coupling and damping rates, it remains open whether bulk TMDCs are materials with strong light-matter coupling under ambient conditions, i.e., if the coupling strength $g$ exceeds their intrinsic damping rate $\gamma$. 

Here we determine light-matter coupling in bulk MoS$_2$ from the polariton dispersion measured in thin slabs. We measured the polariton wavelength with a scattering-type SNOM for various excitation wavelengths in the visible (700-540\,nm). From the fits of the dispersion we obtained a coupling strength of  40\,meV for the $A$ exciton and $100\,$meV for the $B$ exciton at room temperature. Both excitons in MoS$_2$ are in the strong coupling regime with similar coupling strength expected in other TMDCs. TMDCs are excellent materials to exploit polaritons under ambient conditions from three-dimensional crystals down to two-dimensional monolayers.

\section{Theory}

Exciton polaritons  are observed in bulk materials by their reflection, absorption, and luminescence spectra, but such experiments require highly pure samples.\cite{Sell1973,Adreani1995} Alternatively, propagating exciton-polariton modes may be imaged in thin slabs by scanning near-field spectroscopy (SNOM) as suggested by Hu~\textit{et al}.\cite{Hu2017} They used TMDC slabs that were thick enough to prevent quantum confinement of the exciton states, but thin enough to support waveguided photons.\cite{Hu2017a,Khurgin2015} Such measurements  can be used to determine the polariton properties of the bulk material as we will show now. 

To describe polariton formation in a thin TMDC slab, we first introduce the dielectric properties of a bulk exciton polariton and then consider the solutions for the waveguided modes. An exciton polariton forms through the interaction of an exciton $\omega_{ex}$ with a photon $\omega_{pt}$. Inside the material the two independent (quasi)particles, photon and exciton, are replaced by the exciton-polariton as a new quasiparticle. The formation of the coupled state has some interesting consequences; for example, radiative decay is no longer a loss channel for the material excitation, because the polariton continuously converts from a matter into a photonic excitation and vice versa.\cite{Adreani1995} We consider a layered material where the in-plane dielectric function $\varepsilon_i(\omega)$ is given by a background dielectric constant $\varepsilon_b$ plus resonances by two excitons. Such an ansatz describes the contribution of the $A$ and $B$ exciton of MoS$_2$ to the optical properties of the material.\cite{Neville1976,Adreani1995}
\begin{equation}
    c^2 Q^2=\varepsilon_i(\omega)\omega^2=
    \epsilon_b \left( 1-\frac{4g_A^2}{\omega^2 - \omega_{A}^2 + i \gamma_{A} \omega} - \frac{4g_B^2}{\omega^2 - \omega_{B}^2 + i \gamma_{B} \omega} \right)\omega^2.
    \label{EQ:ep}
\end{equation}
Here $\omega_{ex}$ ($ex=A, B$) are the energies of the $A$ and $B$ exciton and $\gamma_{ex}$ their damping constants. $Q$ is the complex exciton-polariton wavevector. The out-of-plane dielectric function of the material $\varepsilon_o(\omega)$ is assumed to be constant.

Equation~\eqref{EQ:ep} connects two complex quantities, namely, frequency $\omega$ and wavevector $Q$, and requires a four-dimensional plot for representation.\cite{Wolff2018} The standard way of visualizing the polariton dispersion is to impose real values for $Q$ and plot the real part of the frequency $Re(\omega)$ versus $Re(Q)$. In this plot the polariton dispersion contains two branches with a minimum separation $\Omega_r^{ex}$. These so-called upper and lower polariton branch are indeed observed in experiments, if the experimental setup imposes real values on the polariton wavevector, e.g., luminescence, light scattering and reflection.\cite{Arakawa1973,Henry1965,Mueller2020,Munkhbat2019} SNOM, in contrast, imposes real frequencies, because the system is driven by a laser and exciton propagation and decay are observed in real space.\cite{Hu2017,Fei2016,Wolff2018} The $Re(\omega)$ over $Re(Q)$ dispersion  then shows a backbending close to the resonance frequency. The Rabi splitting is found by the energetic difference between the lowest and highest polariton energies at the crossing point between the dispersion of light and the exciton resonance energy.\cite{Wolff2018,Arakawa1973}

We now consider electromagnetic waves in a thin slab of a material. In a slab with thickness $d$ photons propagate as quantized waveguide eigenmodes ($d\ll\lambda_{pt}$, $\lambda_{pt}$ vacuum photon wavelength). A free standing slab in air supports a transverse electric (TE) and transverse magnetic (TM) waveguide mode down to vanishing slab thickness,\cite{Khurgin2015,Hu2017a} but a minimum slab thickness is required to guide photons in an asymmetric dielectric environment like a flake on a substrate.\cite{Khurgin2015} We are particularly interested in the thickness range around $d\approx50$\,nm where only the lowest-order TE$_0$ and TM$_0$ modes are allowed.\cite{Hu2017a} The dispersion of the TE$_0$ and TM$_0$ modes are described by effective dielectric functions that depend on the slab thickness and the dielectric properties of the slab and its surrounding.\cite{Hu2017a,Ermolaev2020} The effective dielectric function for the TM$_0$ mode $\varepsilon_\mathrm{TM}$ is found from the condition\cite{Hu2017a}
\begin{equation}
 \frac{2\pi d}{\lambda_{pt}}\sqrt{\varepsilon_i(1-\varepsilon_\mathrm{TM}/\varepsilon_o)}=
 \tan^{-1}\left[
\frac{\varepsilon_i\sqrt{\varepsilon_\mathrm{TM}-1}}{\sqrt{\varepsilon_i(1-\varepsilon_\mathrm{TM}/\varepsilon_o)}} 
\right]
+ 
\tan^{-1}\left[
\frac{\varepsilon_i\sqrt{\varepsilon_\mathrm{TM}-\varepsilon_s}}{\varepsilon_s\sqrt{\varepsilon_i(1-\varepsilon_\mathrm{TM}/\varepsilon_o)}} 
\right].
\label{EQ:TM}
\end{equation}
$\lambda_{pt}$  is the wavelength of light in vacuum and $d$ the thickness of the slab. We assumed a flake in air on a substrate with $\varepsilon_s$ ($\varepsilon_s=2.1$ for SiO$_2$). $\varepsilon_i$ is the dielectric function of the material for light polarized along the planes, see Eq.~\eqref{EQ:ep} and  $\varepsilon_o$ the dielectric function for out-of-plane polarization A similar condition yields the effective dielectric function for the TE$_0$ mode $\varepsilon_\mathrm{TE}$\cite{Hu2017a}
\begin{equation}
   \frac{2\pi d}{\lambda_{pt}}\sqrt{\varepsilon_i-\varepsilon_\mathrm{TE}}= 
 \tan^{-1}\left[
\frac{\sqrt{\varepsilon_\mathrm{TE}-1}}{\sqrt{\varepsilon_i-\varepsilon_\mathrm{TE}}} 
\right]
+    
\tan^{-1}\left[
\frac{\sqrt{\varepsilon_\mathrm{TE}-\varepsilon_s}}{\sqrt{\varepsilon_i-\varepsilon_\mathrm{TE}}} 
\right].
\label{EQ:TE}
\end{equation}

\begin{figure}
    \centering
    \includegraphics[width=8.5cm]{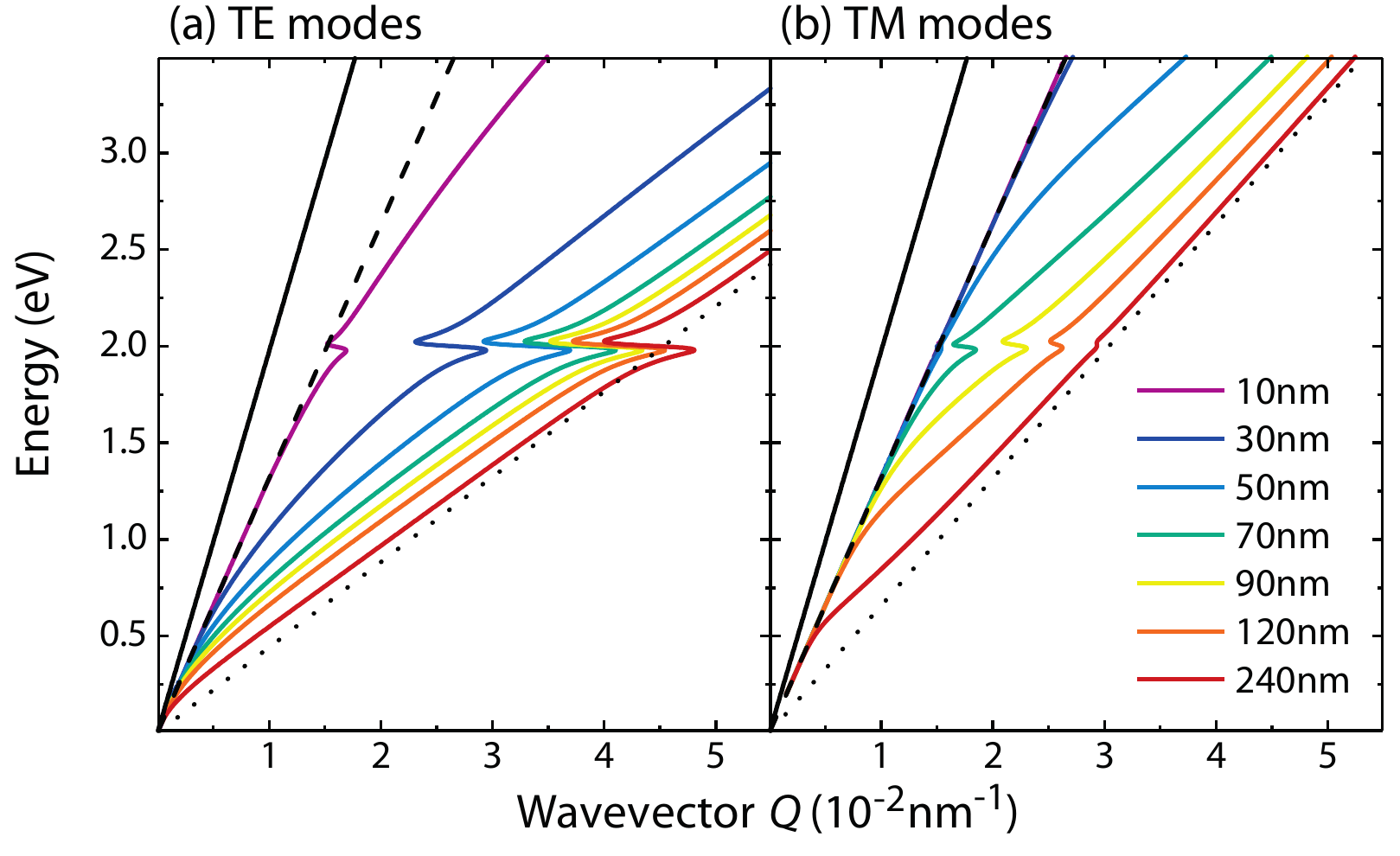}
    \caption{Dispersion of the (a) TE$_0$ and (b) TM$_0$ mode in a thin slab of a layered material with an exciton resonance. The slab thickness is shown in the legend in (b). The full line is the photon dispersion in air, the dashed line the dispersion of the substrate with $\varepsilon_s=2.25$, and the dotted line the dispersion for a dielectric constant (a) $\varepsilon_b=20$ and (b) $\varepsilon_o=9$. The exciton was modeled as being polarized within the plane with $\omega_{ex}=2\,$eV, $\gamma_{ex}=50\,$meV, and $g_{ex}=100\,$meV.}
    \label{FIG:TETM}
\end{figure}

The characteristic dispersion of the TE$_0$ and TM$_0$ mode in slabs of layered material are shown in Fig.~\ref{FIG:TETM}. Since the dispersion of both modes contains features originating from the in-plane dielectric function, Eqs.\eqref{EQ:TM} and \eqref{EQ:TE}, the TE or TM dispersion may be used to determine $\varepsilon_i$ of the bulk material. The TE$_0$ mode exists already for very thin layers (10\,nm), purple line in Fig.~\ref{FIG:TETM}(a). The polariton-induced backbending is only weakly pronounced for thin slabs, because of the limited overlap between the TE$_0$ mode and the slab material. The dispersion of this in-plane polarized mode  rapidly converges to $\varepsilon_i$ with increasing $d$, red and dotted lines in Fig.~\ref{FIG:TETM}. The TE$_0$ mode is independent of the out-of-plane dielectric function $\varepsilon_o$, see Eq.~\ref{EQ:TE}.
The situation is quite different for the TM$_0$ mode, where the electric field is predominantly oriented perpendicular to the layers, but also contains in-plane contributions. This mode requires a minimum thickness of $d=50\,$nm, light blue line in Fig.~\ref{FIG:TETM}(b); for large $d$ it converges to $\varepsilon_o$, dotted line. Similar to the TE$_0$ mode, the backbending of the TM$_0$ mode is small for small $d$. The backbending increases with  $d$, but exists only due to the in-plane component of the electric field, which vanishes for large $d$. The result is a maximum in the backbending for a thickness of 90\,nm in Fig.~\ref{FIG:TETM}(b) and a smaller Rabi splitting than for the TE$_0$ mode at any given thickness. In any case, fitting the TE or  TM modes in thin slabs with Eqs.~\eqref{EQ:TM} and \eqref{EQ:TE}  yields the polariton dispersion or dielectric function of the underlying bulk material.

\section{Experiments}

\begin{figure}
\includegraphics[width=8cm]{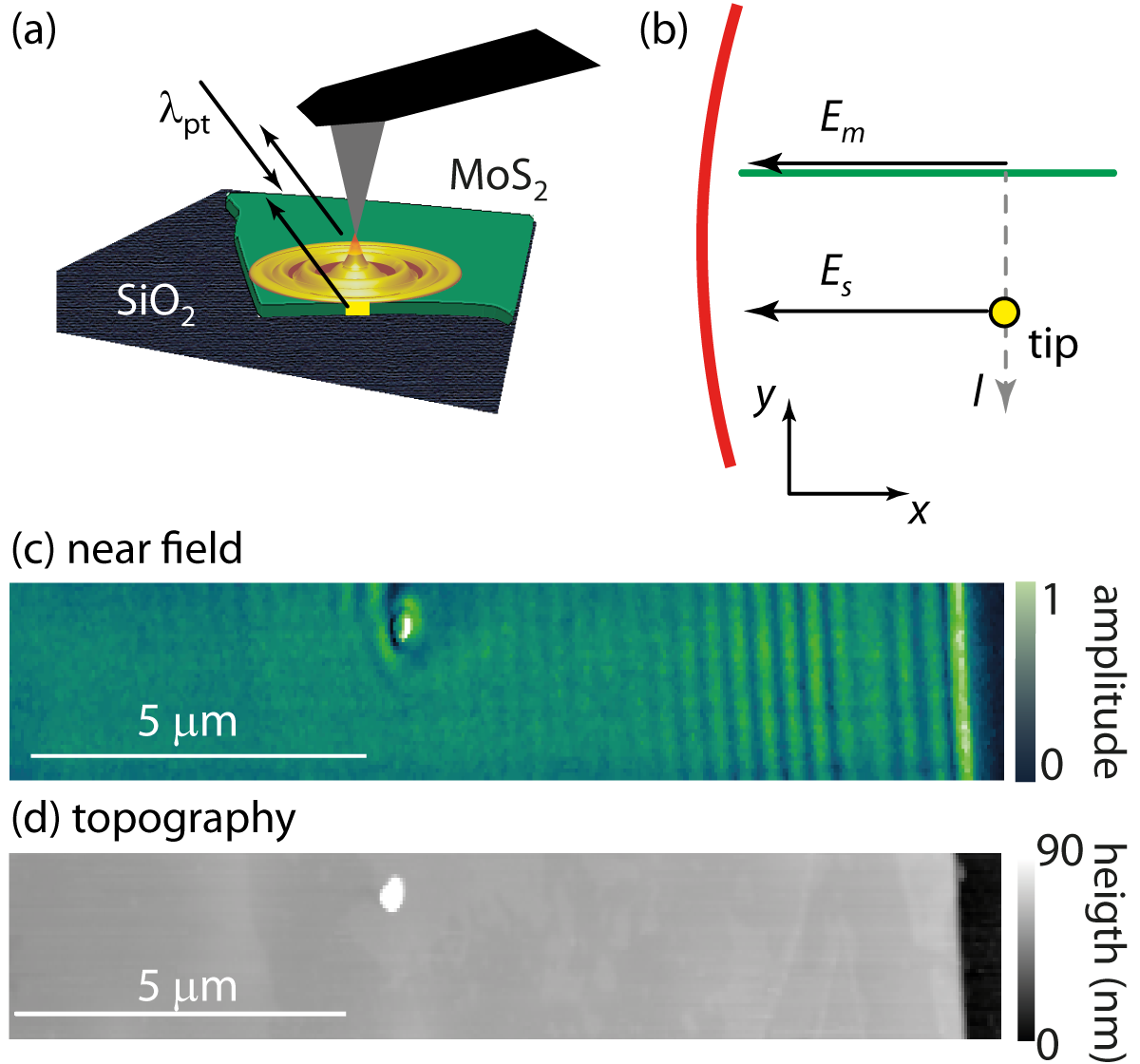}
\caption{Near-field imaging. (a) Sketch of the metallic tip that excites an exciton polariton in a flake of MoS$_2$ (green) when illuminated by a laser with wavelength $\lambda_{pt}$.  The polariton scatters at the edge of the sample and is emitted partly into free space. Light gets scattered also by the tip directly, which will interfere with the edge-emitted photons. (b) On-top view of light emission in the $\theta=0^\circ$ configuration. $\theta$ is the angle between the optical axis of the mirror and the sample edge, see Supplementary Information for details. The electric fields $E_s$ produced at the tip  and $E_m$ produced at the edge interfere in the detection system (indicated as the red line representing the collecting mirror). Yellow dot: tip, green line: MoS$_2$ edge. The inset at the bottom shows the in-plane corrdinate system. (c) Near-field amplitude on an MoS$_2$ flake with $d=68\,$nm excited at $\lambda_{pt}=578\,$nm. Clearly visible are the interference patterns from the superposition of $E_s$ and $E_m$. (d) Topography image recorded simultanesouly with (c).}
\label{FIG:sSNOM}
\end{figure}

To measure the dispersion of the waveguided modes, we imaged their near-fields on MoS$_2$ flakes using a scattering-type SNOM (s-SNOM, Ref.~\cite{Taubner2003}) operating at excitation energies in the visible, Fig~\ref{FIG:sSNOM}. Polaritons manifest in near-field images through  interference between various scattering pathways in the SNOM. To obtain the dispersion, we repeated the experiment for various excitation wavelengths and determined the polariton wavevector as a function of energy. Such experiments have been reported only for infrared wavenlengths or single-line laser excitation.\cite{Hu2017,Hu2017a,Ermolaev2020} Here, we implement it for visible excitation (up to 540\,nm) using a fully tunable, narrow-line, and noise-suppressed laser.

Thin flakes of MoS$_2$ on a Si substrate with 300\,nm SiO$_2$ were prepared by mechanical exfoliation. The substrate was sonicated (20-30\,min), washed (isopropanol, acetone), and dried under nitrogen gas. We cleaved a piece of MoS$_2$ several times with a scotch tape and then transferred it to the freshly prepared substrate. We examined the size and thickness of the flakes by atomic force microscopy (AFM). The SNOM measurements were conducted on flakes of sub-wavelength thickness that are suitable as waveguides. In this paper we report the results obtained on a flakes with $d=58, 68,$ and $82$\,nm.

 In the s-SNOM (neasnom by neaspec) the laser illuminates a metallic AFM tip, Fig.~\ref{FIG:sSNOM}a. The  tip produces an optical near field that excites slab polaritons in the flake. A mirror collects the light that gets scattered by the tip or emitted by the sample, Fig.~\ref{FIG:sSNOM}b. The light is demodulated in a pseudo-heterodyne detection setup (3$^\mathrm{rd}$ harmonic),\cite{Hillenbrand2001} which  
produces a near-field image of the sample, Fig.~\ref{FIG:sSNOM}c, with characteristic fringes and patterns that we will further analyse below. The AFM topography of the same area, Fig.~\ref{FIG:sSNOM}d, is completely flat and featureless.
The microscopy images were produced simultaneously using enhanced platinum tips (neaspec) working in tapping mode at the tip eigenfrequency $\approx 350$\,kHz with an amplitude of 50\,nm. 
 A unique feature of our s-SNOM system is the wide range of laser wavelengths in the visible 700-450\,nm (1.77-2.75\,eV) that are provided by the fully tunable C-Wave laser (H\"ubner Photonics). Similar near-field images as in Fig.~\ref{FIG:sSNOM} were obtained for the energy range 1.79-2.3\,eV, which covers the $A$ and $B$ exciton resonance of MoS$_2$.

There are several mechanisms that can give rise to the interference patterns in Fig.~\ref{FIG:sSNOM}c. The dominant interference mechanism when exciting TMDCs in the visible is that edge emitted light with electric field amplitude $E_{m}$ interferes with light that was scattered directly by the metallic tip $E_s$, Fig.~\ref{FIG:sSNOM}b.\cite{Hu2017,Fei2016} We worked in a configuration where the emmissive edge is parallel to the optical axis of the collecting mirror as sketched in top view in Fig.~\ref{FIG:sSNOM}b ($\theta=0^\circ$, see Supplementary Information). We chose this configuration, because it is free of systematic errors that arise from the limited knowledge of the exact alignment configuration.\cite{Hu2017} A near-field image as in Fig.~\ref{FIG:exp}c shows the field amplitude of the tip and edge emitted waves $A_{nr}\propto|E_s+E_m|$. In the $\theta=0^\circ$ configuration and for a tip position at a point $l$ along a line perpendicular to the edge, $A_{nr}$ at the detector is given by, see Supplementary Information,
\begin{equation}
    A_{nr}(l)\propto|E_s+E_{m}|=\sqrt{A_s^2+A_{m}^2 e^{-2\gamma_Ql}+2A_sA_{m} e^{-\gamma_Ql}\cos{\frac{2\pi}{\lambda_{ep}}l}}
    \label{EQ:fit}
\end{equation}
with
\begin{equation}
    E_s=A_s e^{i(k_x x+\omega t)}
\end{equation}
and
\begin{equation}
    E_{m}=A_{m}e^{i Q l}e^{i(k_x x + \omega t)}.
\end{equation}
$A_s$ is the amplitude of the light scattered by the tip and $A_{m}$ the amplitude of the light emitted at the edge towards the collecting mirror. $Q$ is the complex wavevector of the exciton polariton propagating along the $y$ direction, see Fig.~\ref{FIG:sSNOM}b for the coordinate system. Its real part $\textrm{Re}(Q)=2\pi/\lambda_{ep}$ yields the polariton wavelength $\lambda_{ep}$; its imaginary part $\textrm{Im}(Q)=\gamma_Q$ is the damping of the polariton. $k_x$ is the $x$ component of the photon in free space. 

We extracted line scans perpendicular to the sample edge from the near-field images as in Fig.~\ref{FIG:sSNOM}c. The fit with Eq.~\eqref{EQ:fit} yields the wavelength and damping of the polariton. Equation~\eqref{EQ:fit} is valid only for  $\theta=0^\circ$ as in Fig.~\ref{FIG:sSNOM}b. The fringe period -- i.e., the distance between two interference maxima -- deviates rapidly from $\lambda_{ep}$ for any misalignment. We carefully verified the dependence and corrected the measured $\lambda_{ep}$ for the misalignment, see Supplementary Information for details.

\section{Results}

\begin{figure}
    \centering
    \includegraphics[width=8.5cm]{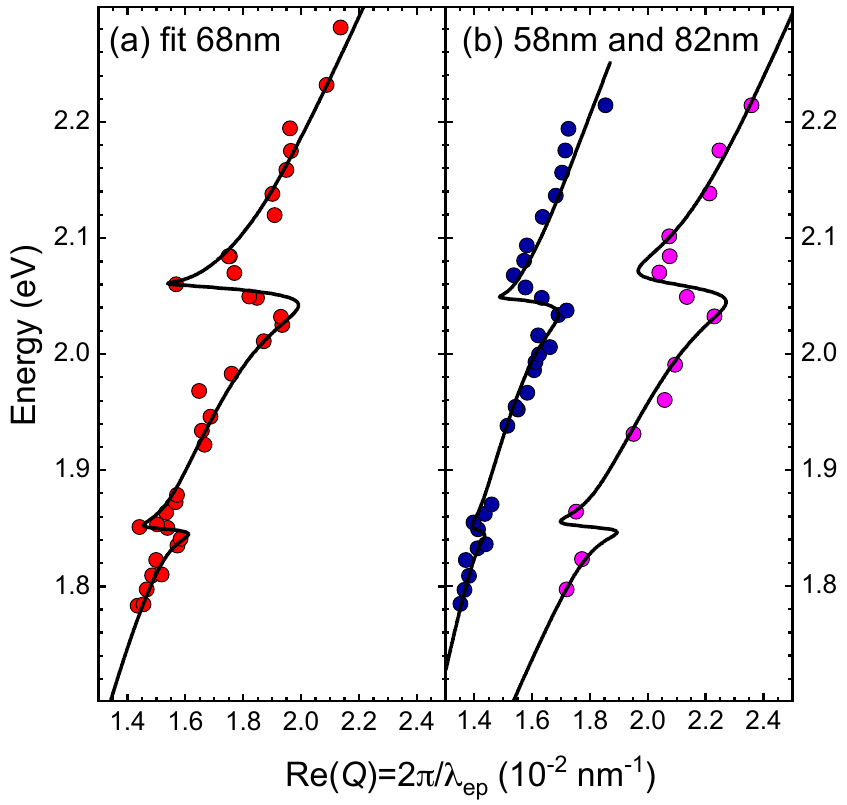}
    \caption{Experimental dielectric function of thin MoS$_2$ slabs. Dots: dispersion of the TM$_0$ mode measured with the s-SNOM at a flake with (a) $d=68$\,nm, (b) 58\,nm (blue), and 82\,nm (purple). (a) The line is a fit to the experimental data with Eq.~\eqref{EQ:TM}. (b) The lines are plots of the predicted TM$_0$ dispersion using the fit parameters of panel (a), see Table~\ref{TAB:fit}.}
    \label{FIG:exp}
\end{figure}

The experimental dispersion of MoS$_2$ flakes with thicknesses below 100\,nm shows two backbending slopes around 1.85 and 2.05\,eV, see Fig.~\ref{FIG:exp}. The backbending is weakest for the thinnest flake with $d=58\,$nm. It increases for 68\,nm and then again gets weaker for the thickest 82\,nm flake. The increase and decrease of the backbending  is characteristic of the TM$_0$ mode, Fig~\ref{FIG:TETM}. We detected no signatures of the TE$_0$ mode in our near-field images. This is reasonable, because the predominantly $z$ polarized near field of the metallic tip cannot couple to TE eigenmodes that are polarized in plane. The line in Fig.~\ref{FIG:exp}a is a fit to the experimental data obtained at the $d=68\,$nm flake with Eq.~\eqref{EQ:TM}. We used the same parameters, Table~\ref{TAB:fit}, to calculate the dispersion for $d=58\,$ and $82\,$nm. The fit describes the data obtained in the other two flakes very well, Fig.~\ref{FIG:exp}b. In particular, it reproduces the decreasing Rabi splitting for smaller and larger thicknesses.

\begin{table}
    \centering
    \begin{tabular}{lccccc}\hline
&$\omega_{ex}$ (eV)&$\gamma_{ex}$ (meV)&$\Omega_r$ (meV)&$g$ (meV)&$\omega_{ex}^{LT}$ (meV)\\\hline
A exciton&1.850&11&84&42&1.6\\
B exciton&2.053&35&194&97&9.4\\\hline
    \end{tabular}
    \caption{Fit parameters of the MoS$_2$ excitons.
    $\omega_{ex}^{LT}$ is the splitting of the transverse and longitudinal exciton related to the Rabi splitting by 
    $\Omega_r^{ex}=\sqrt{2\omega_{ex}^{LT}\omega_{ex}}$
    }
    \label{TAB:fit}
\end{table}

The pronounced backbending of the TM$_0$ dielectric function  implies that the $A$ and $B$ excitons  of the MoS$_2$ slabs are in the regime of strong light-matter coupling.\cite{Baranov2018,Wolff2018} More importantly, the coupling remains strong between excitons in bulk MoS$_2$ and free-space photons, Table~\ref{TAB:fit}. We find a  coupling  strength $g_A=\Omega_r^A/2=42\,$meV for the $A$ exciton in bulk MoS$_2$ that is four times larger than its damping $\gamma_A=11\,$meV. The coupling strength of the $B$ exciton $g_B=97\,$meV  exceeds its decay rate $\gamma_B=35\,$meV almost by a factor of three. The Rabi splitting in MoS$_2$ is an order of magnitude larger than in classical semiconductors like GaAs. It rivals the coupling in found in wide band-gap semiconductors with low dielectric constants,\cite{Sanvitto2016} but the dielectric constants of MoS$_2$ are large. We find an out-of-plane constant $\varepsilon_o=7.6$ and an in-plane background dielectric constant $\varepsilon_b=25$, in good agreement with previous measurements.\cite{Hu2017a,Ermolaev2020} At the same time, the exciton energies $\omega_A=1.85\,$eV and $\omega_B=2.05\,$eV make MoS$_2$ an attractive material for polariton-based devices at visible and near-infrared wavelengths.\cite{Sanvitto2016}

Interestingly, the exciton polaritons of MoS$_2$ also fulfill the condition for very strong light-matter coupling. Very strong coupling means that the coupling strength is comparable to the exciton binding energy $E_{ex}^b$, i.e., $\beta_{ex}=g/E_{ex}^b\approx 1$.\cite{Khurgin2001} To calculate $\beta$ we need the exciton binding energies of bulk MoS$_2$ that, surprisingly, remain debated.\cite{Saigal2016,Neville1976,Fortin1975} Depending on the interpretation of the optical spectra $E_A^b=40-80\,$meV and $E_B^b=100-130\,$meV placing $\beta_A\approx 0.6-1.25$ and $\beta_B\approx 0.8-1$. For $\beta$ approaching one, light-matter coupling contributes to the interaction between the electron and the hole forming the exciton.\cite{Khurgin2001} They start to interact by emitting and absorbing virtual photons, which reduces the exciton radius in the lower and increases the radius in the upper polariton state. The condition for very strong coupling is only met by bulk MoS$_2$ and lost for thin slabs and monolayers, because the coupling decreases to $g\approx 20\,$meV (Ref.~\cite{Liu2014}) and the binding energy increases to $E_b\gtrsim 200\,$meV (Ref.~\cite{Zhang2014,Wang2018}) so that $\beta$ drops to $0.1$ or less.

\begin{figure}
    \centering
    \includegraphics[width=8.5cm]{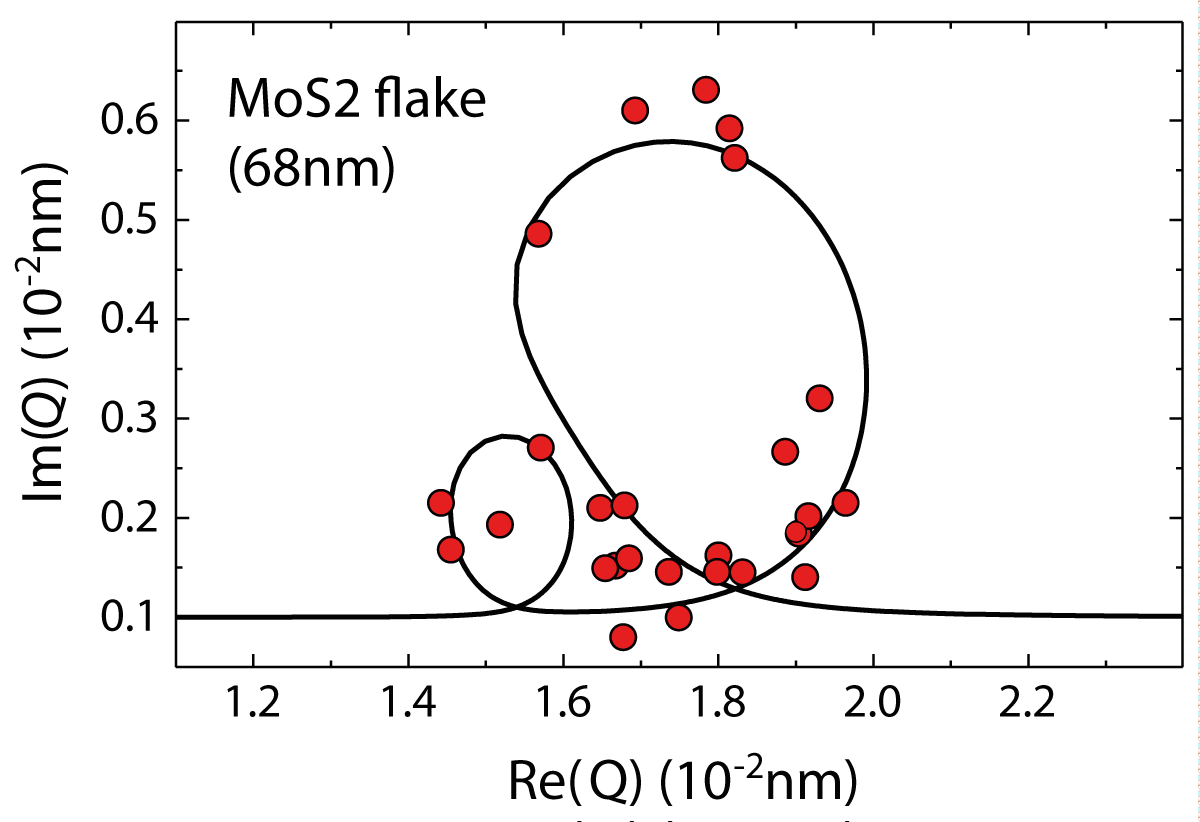}
    \caption{Imaginary and real part of the polariton wavevector in the $d=68\,$nm MoS$_2$ flake. The data points were obtained from the near-field images, see Fig.\,S1. The line is a fit obtained from the data in Fig.~\ref{FIG:exp}a, see text. }
    \label{FIG:Imk}
\end{figure}

The damping rates $\gamma_A$ and $\gamma_B$ extracted from the fits to the TM$_0$ dispersion are smaller than the inhomogeneous linewidth of MoS$_2$ flakes at room temperature.\cite{Mak2010,Wang2018} Nevertheless, they most likely underestimate the lifetime of the MoS$_2$ excitons, because we neglected the damping of the photonic mode via edge emission. $\gamma_B$ is three times larger than $\gamma_A$, which agrees with the range of rates ovwerved for the two excitonic states.\cite{Wang2018PCC,Cha2016} We can also determine the spatial damping of the MoS$_2$ polaritons from the near-field images, because the interference patterns exist over several micrometers, Fig.~\ref{FIG:sSNOM}c. The damping depends strongly on the exciting laser frequency: While the pattern extends for  some micrometers for frequencies away from the $A$ and $B$ exciton, Supplementary Fig.~S1, it drops to  hundreds of nanometers for resonant excitation. We extract the spatial damping from the fits with Eq.~\eqref{EQ:fit}. A plot of the imaginary versus the real part of the wavevector $Q$ in Fig.~\ref{FIG:Imk} shows two loops at the wavevector of the excitons in the uncoupled system. The looping shape is another signature of strong coupling in photonic systems confirming the result of the $Re(\omega)$ versus $Re(Q)$ plot in Fig.~\ref{FIG:exp}.\cite{Wolff2018} We calculated the full line in Fig.~\ref{FIG:Imk} using Eq.~\eqref{EQ:TM} with the fitting parameters obtained from the polariton dispersion and adding a constant damping of $10^{-3}$\,nm$^{-1}$. The latter represents scattering by crystal imperfections and errors introduced by the experimental setup. With increasing distance between tip and edge the edge emitted photons propagate off the optical axis, which eventually reduces the interference amplitude. The maximum propagation length $l_p=1/2Im(Q)$ of 5\,$\mu$m in Fig.~\ref{FIG:Imk}, therefore, poses a lower bound to the polariton propagation. In resonance with the $B$ exciton the propagation length drops to 100\,nm, but even this value is longer than for the bulk polariton. Using the parameters of Table~\ref{TAB:fit} we predict a minimum propagation length $\approx 20\,$nm for polaritons at the exciton energies in the bulk. This high damping is a consequence of the strong coupling, because these polariton frequencies correspond to the forbidden gap that opens between lower and upper polariton in the dispersion for real $Q$ vectors.\cite{Adreani1995,Wolff2018} The propagation length far away from resonances in bulk MoS$_2$ is overestimated in our experiments; it should  be evaluated in crystals of high quality at low temperatures.

\begin{figure}
    \centering
    \includegraphics[width=8.5cm]{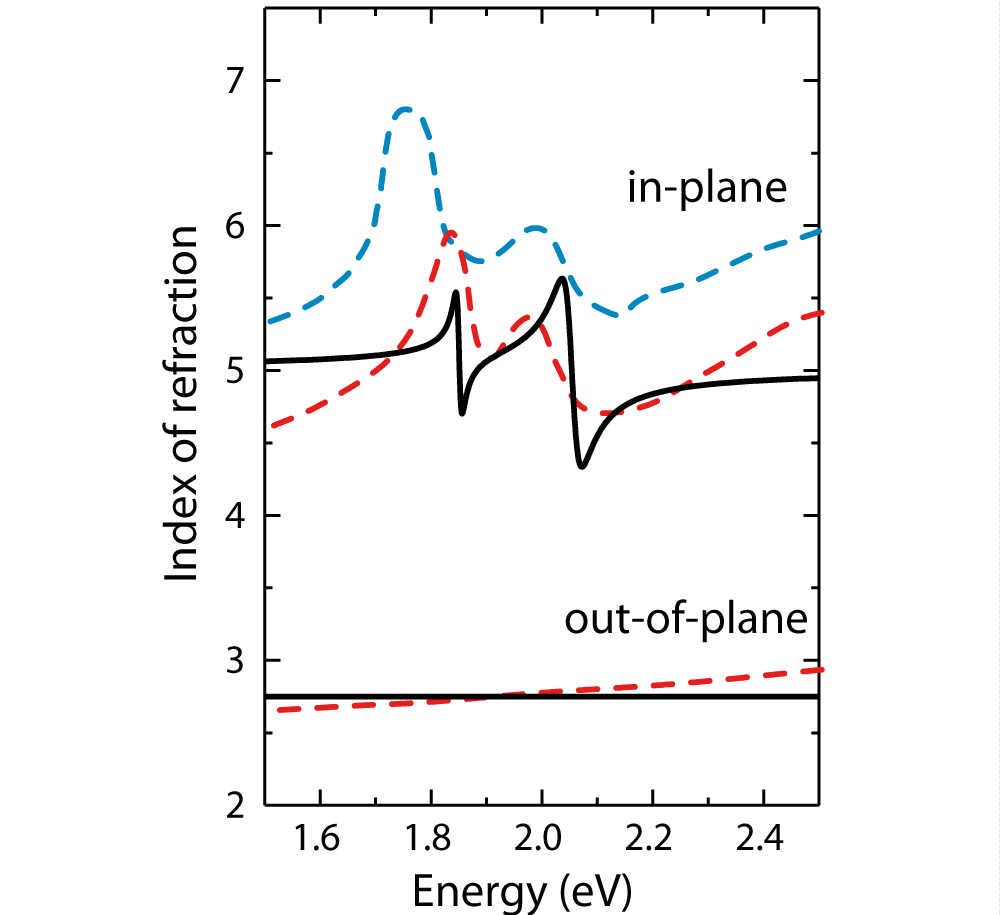}
    \caption{Index of refraction in MoS$_2$. The full line shows the polaritonic contribution to the index of refraction as determined from the slab measurements. The dashed red line is for ellipsometry measurements (Ref.~\cite{Ermolaev2020}) and the dashed blue line from optical reflectance (Ref.~\cite{Neville1976}).}
    \label{FIG:n}
\end{figure}

It is instructive to deduce the polariton contribution to the bulk dielectric function in MoS$_2$ from our slab measurements. Figure~\ref{FIG:n} shows the polaritonic part of the index of refraction compared to reflectance (blue) and ellipsometric (red) measurements on bulk MoS$_2$.\cite{Neville1976,Ermolaev2020} The overall agreement is quite remarkable confirming that the optical response of MoS$_2$ in the visible is dominated by exciton-related effects. The bulk measurements are much broader in line width than the polaritonic part of the index of refraction. This is due to band to band transitions and defect-related excitations. It would be interesting to obtain more experimental data at low temperatures to better distinguish between intrinsic and extrinsic sources of damping. The similarity between the polaritonic contribution to the index of refraction obtained from the slabs and the experiments on bulk MoS$_2$ show that polaritons need to be considered in the description of this material. We note that the refractive indices in Fig.~\ref{FIG:n} are strongly anisotropic with a difference  $\Delta n_{oi}\approx 2$ between the in-plane and out-of-plane refractive index.\cite{Neville1976,Ermolaev2020} Polaritons can be guided and confined efficiently along the MoS$_2$ layers even in the bulk material.

\section{Conclusion}

In conclusion, we measured the dispersion of the transverse magnetic modes in slabs of MoS$_2$ using near-field optical microscopy with fully tunable visible excitation. We  determined the strength of light-matter coupling in bulk MoS$_2$ from the dispersion of the slab mode. The $A$ and $B$ exciton are in the regime of strong light-matter coupling with a coupling strength of 40\,meV for the $A$ and 100\,meV for the $B$ exciton, which exceeds their losses by more than a factor of two. Because their coupling strength is also comparable to the exciton binding energy, the formation of the exciton and its properties depend also on the coupling to light. MoS$_2$ combines strong light-matter coupling with a large index of refraction and a strong anisotropy in its dielectric response. It is a very promising material for polariton-based devices operating under ambient conditions.

\section{Acknowledgements}

We acknowledge financial support by the European Research Council (ERC) under grant DarkSERS (772108) and by the German Science Foundation (DFG) within the Priority Program SPP 2244 “2DMP”.

\bibliography{LM_MoS2}

\end{document}